\documentclass[12pt,amsfonts]{revtex4}
\usepackage{amsfonts}
\usepackage{amsmath}
\usepackage{bm}
\usepackage{epsf}

\newcommand{\beq}{\begin{equation}}
\newcommand{\eeq}{\end{equation}}
\newcommand{\beqa}{\begin{eqnarray}}
\newcommand{\eeqa}{\end{eqnarray}}

\newcommand{\ket}[1]{| #1    \rangle }

\newcommand{\ave}[1]{  \langle #1   \rangle }

\newcommand{\rref}[1]{~(\ref{#1})}

\begin{document}

%%%%%%%%%%%%%%%%%%%%%%%%%%%%%%%%%%%%%%%%%%%%%%%%%%%%%%%%%%%%%%%%%%%%%%%%
%%%%%%%%%%%%%%%%%%%% Local Definitions %%%%%%%%%%%%%%%%%%%%%%%%%%%%%%%%%

\newcommand{\suba}[1]{_{_{A_{#1}}}}
\newcommand{\subat}[1]{_{_{\widetilde{A}_{#1}}}}
\newcommand{\subb}[1]{_{_{B_{#1}}}}
\newcommand{\subbt}[1]{_{_{\widetilde{B}_{#1}}}}
\newcommand{\subab}[1]{_{_{A_{#1}B_{#1}}}}
\newcommand{\subabt}[1]{_{_{\widetilde{A}_{#1}\widetilde{B}_{#1}}}}

\newcommand{\diracsl}[1]{\slash{\!\!\!#1}}

%%%%%%%%%%%%%%%%%%%%%%%%%%%%%%%%%%%%%%%%%%%%%%%%%%%%%%%%%%%%%%%%%%%%%%%%
%%%%%%%%%%%%%%%%%%%%%%%%%%%%%%%%%%%%%%%%%%%%%%%%%%%%%%%%%%%%%%%%%%%%%%%%

\title[Modewise Entanglement]
{BCS-like Modewise Entanglement of Fermion  Gaussian States  }

\author{ Alonso Botero }
\email{abotero@uniandes.edu.co} \affiliation{
    Departamento de F\'{\i}sica,
    Universidad de Los Andes,
    Apartado A\'ereo 4976,
    Bogot\'a, Colombia}
\author{ Benni Reznik }
\email{reznik@post.tau.ac.il} \affiliation{ Department of Physics
and Astronomy, Tel-Aviv University, Tel Aviv 69978, Israel.
       }

\date{\today}

\begin{abstract}
\bigskip
We show that with respect to any bipartite division of modes, pure
fermion gaussian states display the same type of structure in its
entanglement  of modes as that of  the  BCS wave function, namely,
that of a tensor product  of entangled two-mode squeezed fermion
states. We  show that this structure applies to a wider class of
``isotropic"  mixed fermion states, for which we derive  necessary
and sufficient conditions for mode entanglement.

\end{abstract}

\maketitle

The nature of many-body entanglement in various solid-state models
such as spin chains, superconductivity, and harmonic chains, has
been a recent subject of intense investigation [1-12]. The
motivation for this effort is two-fold: on the one hand, such
systems exhibit  a rich entanglement structure which could
potentially be harnessed for quantum information processing
\cite{divincenzo,Loss,beenakker,verstraete2}; on the other hand,
this effort promises to deepen our  understanding of certain
universal features in many-body physics--such as quantum phase
transitions-- and their relation to entanglement [2-5,12].

In this Letter we investigate the properties of bipartite fermion
mode entanglement \cite{zanardi} for a wide class of models
describing correlated fermion systems, and  show that this
entanglement displays a simple, universal structure. As it is well
known, under certain field transformations or after suitable
approximations, a large class of interacting theories can be
mapped to effective theories described by quadratic Hamiltonians
of the generic form
\begin{equation}\label{hamform}
H = \sum C_{i j} b_i^\dagger b_j + \sum (A_{i j}b_i^\dagger
b_j^\dagger + \mathrm{h.c.}) \, ,
\end{equation}
which are then diagonalized through Bogoliubov-Valatin (i.e.,
canonical) transformations\cite{BogoVal} to  appropriate
quasi-particle mode bases. Important classes of such theories
include the Hubbard Model and the BCS theory of superconductivity
in the 'Mean Field' or Hartree-Fock approximation, as well as
certain exactly solvable spin-chain models after a Jordan-Wigner
transformation \cite{martin,fradkin}. As we shall shortly clarify,
an important feature of the above class of Hamiltonians is that
not only the ground states (the quasi-particle vacuum), but also
all eigenstates describing a definite set of quasi-particle
excitations, belong to the general class of \emph{ fermion
gaussian states}. The bipartite entanglement structure of the
eigenstates of \rref{hamform} is then  given by the following
result, which is applicable to  all pure gaussian fermion states,
and which constitutes the main result of this Letter:

{\it Given a collection of $N$-fermion systems or ``modes",
partitioned into two arbitrary sets,  $A = \{A_i,\ldots,A_m\}$ and
$B = \{B_{1}\ldots,B_n\}$, of sizes $m$ and $n=N-m$, respectively,
any  pure fermion gaussian  state $\ket{\psi}_{AB}$ of the modes
may always be written as }
\begin{equation}\label{decomp}
\ket{\psi}_{AB} =
\ket{\widetilde{\psi}_1}\subabt{1}\ket{\widetilde{\psi}_2}\subabt{2}\ldots
\ket{\widetilde{\psi}_s}\subabt{s}\ket{0}\subat{F}\ket{0}\subbt{F}
\end{equation}
Here, $s \leq \min(m,n)$, and
$\widetilde{A}=\{\widetilde{A}_1\ldots,\widetilde{A}_m\}$,
$\widetilde{B}=\{ \widetilde{B}_1,\ldots,\widetilde{B}_n\}$ are
new sets of modes obtained from $A$ and $B$, respectively, through
local fermion canonical transformations. The states
$\ket{\widetilde{\psi}_k}$ are two-mode fermion squeezed states of
the form
\begin{equation}\label{tmss}
\ket{\widetilde{\psi}_k}\subabt{k} =
\cos\theta_k\ket{00}\subabt{k} - \sin\theta_k \ket{11}\subabt{k}\,
,
\end{equation}
 entangling the modes $\widetilde{A}_k$ and $\widetilde{B}_k$
for $ k\leq s$, and $\ket{0}\subat{F}$ and $\ket{0}\subbt{F} $ are
products of vacuum states for the remaining modes in
$\widetilde{A}$ and $\widetilde{B}$ respectively.

Together with a similar decomposition obtained for boson gaussian
states in \cite{modewise,giedke}, the above result shows that the
bipartite entanglement structure  of pure multi-mode gaussian
states is of  $1 \times 1$-entangled mode pairs independently of
the statistics, the partition of modes, or  the nature of the
basis in which the partition is performed. The present result
shows that in the fermion case, the mode-pairing is similar to
that of the BCS wave function $\ket{ \mathrm{BCS}} = \Pi_{k}(\cos
\theta_k + \sin \theta_k b_{k\uparrow} ^\dagger
b_{-k\downarrow}^\dagger)\ket{\mathrm{vac}}$, except that the
vacuum state and the pairwise excitations in \rref{decomp} are all
relative to a quasi- particle spectrum determined by  the
multi-mode state $\ket{\psi}$ and the choice of partition.

As in the boson case investigated in \cite{modewise}, it can be
shown that the above theorem and the decomposition \rref{decomp},
follows for an arbitrary pure Gaussian Fermi state from the
properties of the Schmidt decomposition, and from the comparison
of the reduced density matrices in their normal form, at regions A
and B. In this letter we shall, however, follow a more general
framework, which enables us to generalize this theorem to a
certain class of ``isotropic" mixed gaussian states, characterized
by a symmetry property of the covariance matrices.  The remainder
of the Letter is thus structured as follows: we first review
fermion gaussian states, canonical transformations, and define the
fermion covariance matrix; next, we define isotropic gaussian
states and prove a general decomposition for isotropic states in
the language of covariance matrices; finally we also derive a
necessary and sufficient conditions for the entanglement of
isotropic mixed states and close with some remarks on the
applicability of our results.

Let a system of $N$ fermion modes be described by a set of
creation and annihilation  operators $b_i^\dagger, b_i$ satisfying
anticommutation relations $\{ b_i^\dagger  , b_j^\dagger \} = \{
b_i  , b_j \} =0$, $ \{ b_i  , b_j^\dagger \} =\delta_{ij}$. We
define a fermion gaussian state  for such a system as any state
$\rho$ that, with a certain choice of mode basis $ \widetilde{b}_i
= u_{i}{}^{j} b_{i} + v_{i}{}^{j} b^\dagger_{j} \, $ preserving
the fermion anticommutation algebra,  acquires the form
\cite{peschel01}
\begin{equation}\label{rhogauss}
\rho = \bigotimes_{k=1}^{N} \widetilde{\rho}_k \, , \ \ \ \
\widetilde{\rho}_k = \frac{1}{2}\left(1 - \lambda_i[
\widetilde{b}_i^\dagger,\widetilde{b}_i]\right)\, ,
\end{equation}
with $|\lambda_i|\leq 1$ and $|\lambda_i|= 1$ for pure states
(note that $\rho$ can also be written in the form $\rho = Z
^{-1}\exp({-\beta \widetilde{b}_i^\dagger \widetilde{b}_i})$).
Fermion gaussian states share with their boson counterparts the
property that correlation functions for the creation/annihilation
operators are completely determined by the two-point functions
according to Wick's theorem\cite{gaudin}. Moreover, since this
property   is extensible to correlation function pertaining to a
reduced subset of the modes, it follows that any partial (reduced)
density matrix obtained from $\rho$ remains gaussian
\cite{peschel02}.

The gaussian nature of a state is preserved under any unitary
transformation that induces a  canonical linear transformation  of
the fermion variables, in other words, that preserves the
anticommutation relations. To determine the group  of canonical
transformations, it becomes convenient to replace the $N$ creation
and annihilation operators by $2N$ hermitian ``fermion quadrature"
combinations
\begin{equation}
\gamma_{2i-1} = b_i + b_i^\dagger \, , \ \ \ \ \ \gamma_{2i} = i
(b_i - b_i^\dagger) \, .
\end{equation}
The anticommutation relations can then be seen to be a consequence
of an $\mathbb{R}^{2N}$ Clifford  (or Dirac)
algebra\cite{Clifford}
\begin{equation}\label{hermanticoms}
\{ \gamma_\alpha, \gamma_\beta \} = 2\delta_{\alpha \beta} \,
\end{equation}
satisfied by the $\gamma_{\alpha}$. Inspection of
\rref{hermanticoms} then shows that any linear transformation of
the form $ \widetilde{\gamma}_\alpha = O_{\alpha \beta}
\gamma_{\beta} \, , $ where $O \in O(2N)$ (the full orthogonal
group in $2 N$ dimensions), preserves the Clifford algebra
\rref{hermanticoms} and hence the anticommutation algebra of the
fermion fields.

That the   group $O(2N)$  of fermion canonical transformations can
be implemented unitarily on the Fock space (i.e. $ \gamma_\alpha'
= U^\dagger \gamma_\alpha U$),  follows from the following: for
any real vector $v \in \mathbb{R}^{2 N}$, define $\diracsl{v} =
\diracsl{v}^\dagger = v \cdot \gamma $, and $|v|^2 = v \cdot v$
($=\diracsl{v}^2 $) with the usual inner product.  We then verify
that any real vector $a$ with $|a|=1$ defines a unitary operator
$U = \diracsl{a}$, the action of which is an inversion
($\det(O)=-1$) in the orthogonal subspace to $a$
\begin{equation}
\diracsl{a}^\dagger\gamma_\mu\diracsl{a}= [2a_\mu a_\nu
-\delta_{\mu \nu}]\gamma_\nu \, ;
\end{equation}
  in $\mathbb{R}^{2
 n}$, this inversion is equivalent  to an $SO(2N)$ rotation
 times a reflection along $a$.  Since  any orthogonal transformation
can be generated from  a  product of a certain number of such
reflections along different axes\cite{Benn}, it follows that
 any orthogonal transformation of the $\gamma$'s can be
induced by a suitably chosen unitary operator of the form $\,
 U =
\diracsl{a}_1\diracsl{a}_2\diracsl{a}_3\, \dots \diracsl{a}_p \, $
with $|a_k|=1$.

The  group of operations thus defined is the $ Pin(2N)$ group, the
double cover of $O(2N)$\cite{Clifford,Benn,DeWitt}. In turn, the
subgroup of $Pin(2N)$ generated by an even number of reflections
is the covering group of $SO(2N)$, $Spin(2N)$,  associated with
the linear fermion canonical transformations (also known as the
Bogoliubov-Valatin\cite{BogoVal} or squeeze
transformations\cite{Svozil}). More generally, $Pin(2N)$ includes
 improper orthogonal transformations ($\det(O)=-1$)  generated
by an odd number of reflections, and which may in fact be
forbidden due to fermion number superselection rules in a
fundamental fermion theory. This fact, however, does not affect
the generality of the results presented here. Since an improper
transformation is a $SO(2N)$ rotation times some canonical
reflection, an improper transformation is equivalent to a proper
transformation but with the role of the creation and annihilation
operators  exchanged for at most one mode.

An important consequence of the $O(2N)$ equivalence of fermion
gaussian states is that all multiparticle eigenstates obtained
from a given vacuum  are  gaussian. Indeed, if
$\ket{\mathrm{vac}}$ is the ground state of a certain Hamiltonian,
annihilated by the $b_i$ operators in a given quasi-particle
basis, then for each mode it follows that $b_i^\dagger
\ket{\mathrm{vac}} = \gamma_{2 i}\ket{\mathrm{vac}}$;  it
therefore follows that multiparticle states are obtained from some
$U \in Pin(2N)$ acting on $\ket{\mathrm{vac}}$.
 Since the vacuum state is gaussian, the resulting state will also
 be gaussian.

We next define for any  gaussian  quantum state  the fermion
Covariance Matrix (FCM) as
\begin{equation}
M_{\alpha \beta} = \frac{1}{2 i }\left\langle \, [\gamma_{\alpha},
\gamma_{\beta}  \, ] \right \rangle_\rho  \, .
\end{equation}
Being an antisymmetric $2N \times 2N$ matrix, the FCM can always
be brought to the block diagonal form
\begin{equation}\label{canform}
W = O M O^T = \bigoplus_{i=1}^{N} \lambda_i J_2\, , \ \ \ \ J_2 \equiv \left( \begin{array}{cc} 0 & -1 \\
1 & 0  \end{array} \right)\, ,
\end{equation}
by means of a $ O(2N)$ transformation, and with the $\lambda_i
\geq 0$.  In terms of the creation/ annihilation operators
obtained from the transformed gamma's $\widetilde{\gamma} = O
\gamma$, the   gaussian state $\rho$  then takes the form
\rref{rhogauss} with $W$ as its FCM. By setting all the
$\lambda_i$ positive  or zero, we obtain a unique characterization
of  a fermion gaussian state (up to relabelling of the modes)
analogous to the so-called  Williamson normal form of the boson
case \cite{modewise,williamson}. We shall therefore term   the
canonical form $W$ as the fermion Williamson form, and   the $\{
\lambda_i \}$ ($\lambda_i \geq 0$) in $W$ as the Wiliamson
eigenvalues.

Now,  the Williamson spectrum can be obtained from the doubly
degenerate spectrum of the matrix $-M^2$. This provides a simple
characterization of pure gaussian states in terms of their FCM;
since such states are unitarily equivalent to pure states of the
form \rref{rhogauss}, with $\lambda_i = 1\, , \forall i$, the pure
FCM s  satisfy the condition $-M^2 = \openone$. Pure gaussian
states can then be viewed as the extremal states in a more general
class of \emph{isotropic} gaussian states, for which the square of
their FCM's remains invariant under any $O(2N)$ transformation:
\begin{equation}
M^2 = -\lambda_0^2\openone \, , \ \ \ \ \lambda_0 \leq 1 \, .
\end{equation}
It is then possible to show that  if $M$ is the FCM of an
isotropic gaussian state, and if the  modes are separated into two
sectors $A$ and $B$ ($\gamma =\gamma_A \oplus \gamma_B$), then
there exist \emph{local} orthogonal transformations
$\widetilde{\gamma}_A = O_A \gamma_A$ and $\widetilde{\gamma}_B =
O_B \gamma_B$ such that the upon re-ordering the modes the FCM
takes the form
\begin{equation}\label{isoform1}
\widetilde{M} =
\widetilde{M}_{\subabt{1}}\!\oplus\,\widetilde{M}_{\subabt{2}}\oplus\,
...\,\oplus\widetilde{M}_{\subabt{s}}\oplus\widetilde{M}_{\subat{F}}\oplus\widetilde{M}_{\subbt{F}}
\, , \ \ \  \
\end{equation}
 where
$\widetilde{M}_{\subabt{i}}$ is  isotropic   for the entangled
pair  sector $\gamma\subabt{i} =
\gamma\subat{i}\oplus\gamma\subbt{i}$ and of  the form
\begin{equation}\label{isoform2}
\widetilde{M}_{\subabt{i}} = \left(\begin{array}{cccc}
  0 & -\lambda_i &   0& \kappa_i \\
  \lambda_i  & 0 &  \kappa_i & 0 \\
     0& -\kappa_i & 0   & -\lambda_i \\
  -\kappa_i &0  & \lambda_i  & 0\\
\end{array}%
\right) \, , \ \ \ \ \kappa_i^2 + \lambda_i^2=\lambda_0^2\, ,
\end{equation}
and $\widetilde{M}_{\subat{F}}$  and $\widetilde{M}_{\subbt{F}}$
are isotropic FCM's with Williamson eigenvalue $\lambda_0$ for the
remaining  modes in the $A$ and $B$ sectors respectively. From the
correspondence between FCMs and gaussian mixed states, it then
follows  that
  in  the new mode basis, the  isotropic  fermion gaussian state
$\rho^{(0)}_{AB}$ takes the form
\begin{equation}\label{decomp2}
\rho^{(0)} =
\widetilde{\rho}\subabt{1}\otimes\widetilde{\rho}\subabt{2}\otimes...
\otimes\widetilde{\rho}\subabt{s}\otimes\widetilde{\rho}^{(0)}\subat{F}
\otimes\widetilde{\rho}^{(0)}\subbt{F} \, ,
\end{equation}
 where $\widetilde{\rho}\subabt{i}$
are two-mode fermion squeezed states with FCM of the form
\rref{isoform2},
 and $\widetilde{\rho}^{(0)}\subat{F}$ and
$\widetilde{\rho}^{(0)}\subbt{F}$ are local isotropic states for
the remaining modes in $\widetilde{A}$ and $\widetilde{B}$. We
note that for an FCM of the form \rref{isoform2}, the
corresponding two-modes squeezed state may be written as
\begin{equation}\label{twomodemixed}
\widetilde{\rho}\subabt{i} = \frac{1}{4} T_i \left(1\!-\!
\lambda_0[
\widetilde{b}\subat{i}^\dagger,\widetilde{b}\subat{i}]\right)\left(1
\!-\!\lambda_0[
\widetilde{b}\subbt{i}^\dagger,\widetilde{b}\subbt{i}]\right)T_i^\dagger
\end{equation}
with $T_i$ an entangling unitary operator
\begin{equation}\label{ti}
T_i =
\exp\left[\!-\!\left(\widetilde{b}\subat{i}^\dagger\!\widetilde{b}\subbt{i}^\dagger\!
\!+\!\widetilde{b}\subat{i}\!\widetilde{b}\subbt{i}\right)\theta_i
\right] \, , \ \ \ \ \tan 2 \theta_i =  \frac{\kappa_i
}{\lambda_i}\, .
\end{equation}
Thus, in the pure case   $\lambda_0=1$, the two-mode entangled
states become projection operators onto  two-mode squeezed states
$T_i \ket{00}\subabt{i}$  which expand out to the form
\rref{decomp}, whereas the unentangled states become projection
operators onto the vacuum states of the remaining modes.

To prove the decomposition \rref{isoform1},  first perform local
orthogonal transformations $\widetilde{\gamma}_A\oplus
\widetilde{\gamma}_B= (O_A\oplus O_B)\gamma_A \oplus\gamma_B$
bringing each of the local FCMs $M_{A}=\frac{1}{2
i}\ave{[\gamma_A\, , \gamma_A^T]}$ , $M_{B}=\frac{1}{2
i}\ave{[\gamma_B\, , \gamma_B^T]}$ into the canonical Williamson
form \rref{canform}, with all $\lambda_i \geq 0$. The total FCM
thus obtained may be written as
\begin{equation}\label{totcm}
    \widetilde{M}=\frac{1}{2 i}\ave{[\widetilde{\gamma}\, , \widetilde{\gamma}^T]} =\left(%
\begin{array}{cc}
  W_A & \widetilde{K} \\
-\widetilde{K}^T & W_B \\
\end{array}%
\right)\, ,
\end{equation}
with $W_A = \bigoplus_{i=1}^{m}\lambda\subat{i}J_2 $ and $W_B =
\bigoplus_{i=1}^{n}\lambda\subbt{i}J_2$. Substituting into the
definition of the isotropic matrix $M^2 = -\lambda_0^2$, the
following equation  is obtained from \rref{totcm}
\begin{equation}
W_A \widetilde{K}  +  \widetilde{K}  W_B   = 0 \label{rel1c}\,
\end{equation}
Consider then a $2 \times 2$ sub-block $\widetilde{K}_{ij} \equiv
-i \ave{\gamma\subat{i} \gamma\subbt{j}^T}$ of $\widetilde{K}$
connecting the modes with eigenvalues $\lambda\subat{i}$ and
$\lambda\subbt{j}$. From equation \rref{rel1c} and using $J_2^2 =
- \openone_2$,  we find that
\begin{equation}
\lambda\subat{i}\widetilde{K}_{ij} =\lambda\subbt{j} J_2
\widetilde{K}_{ij} J_2 \, .
\end{equation}
For $\lambda\subat{i} \neq \lambda\subbt{j}$ this equation has no
solution other than $\widetilde{K}_{ij}=0$, meaning that modes in
$A$ and modes $B$ with different Williamson eigenvalues decouple.

Thus, let $\widetilde{\gamma}\subat{\lambda}$ and
$\widetilde{\gamma}\subbt{\lambda}$ stand for the  modes in $A$
and $B$ with the same local Williamson eigenvalue $\lambda$, and
group the modes according to their eigenvalues so that
$\widetilde{M}$ takes the Jordan form $\widetilde{M} =
\bigoplus_\lambda \widetilde{M}_\lambda$ where each block
$\widetilde{M}_\lambda$ is the FCM for the modes in $A$ and $B$
with a common local Williamson eigenvalue $\lambda$. Concentrating
on a given $\lambda$, let $g_A$ and $g_B$ be the degeneracies of
$\lambda$ in the Williamson spectra of $W_A$ and $W_B$
respectively, so that $\widetilde{M}_\lambda$ may be written as
\begin{equation}\label{degcm}
    \widetilde{M}_\lambda=\left(%
\begin{array}{cc}
  \lambda J_{2g_A} & \widetilde{K}_\lambda \\
-\widetilde{K}_\lambda^T & \lambda J_{2g_B}\\
\end{array}%
\right)\, , \ \ \ \ J_{2g}\equiv \bigoplus_{i=1}^{g} J_2 \, ,
\end{equation}
Now define $\kappa_\lambda = \sqrt{\lambda_0^2\!-\!\lambda^2}$ and
note that $\widetilde{M}_\lambda$ is also an isotropic FCM with
symplectic eigenvalue $\lambda_0$. Thus, $\widetilde{M}^2 =
-\lambda_0^2$ yields
\begin{subequations}\label{rel2}
\begin{eqnarray}
& \widetilde{K}_\lambda  \widetilde{K}_\lambda^T =
\kappa_\lambda^2\openone_{2 g_A}\, , \ \ \ \
 \widetilde{K}_\lambda^T  \widetilde{K}_\lambda
=  \kappa_\lambda^2
\openone_{2 g_B} &\label{rel2b} \\
&J_{2g_A} \widetilde{K}_\lambda J_{2g_B} = \widetilde{K}_\lambda\,
, & \label{rel2c}
\end{eqnarray}
\end{subequations}
where we have used the fact that  $J_{2g}^2 = -\openone_{2 g}$.
Taking the trace of both equations in \rref{rel2b} and using
$\textrm{Tr}_A[ K_\lambda K_\lambda^T ] = \textrm{Tr}_B[
K_\lambda^T  K_\lambda ]$, we find that $ \kappa_\lambda^2(g_A -
g_B) = 0  $, proving
 that for $\lambda \neq \lambda_0$, the degeneracies of
 $\lambda$ in the local FCM's are the same. Thus, let $g_A =
g_B = g$ and express $\widetilde{K}$ in terms of an unspecified
matrix $Q_\lambda$ according to
\begin{equation}
\widetilde{K}_\lambda \equiv \kappa\, Q_\lambda \beta\, , \ \ \ \ \ \beta \equiv \bigoplus_{i=1}^{N}  \left( \begin{array}{cc} 0 &  1 \\
 1 & 0  \end{array} \right)\, .
\end{equation}
Noting that  $\beta^2 = \openone_{2g}$ and $\{\beta,J_{2g}\} = 0$,
we find from  Eqs.   \rref{rel2} that $Q_\lambda$ satisfies
\begin{equation}
Q_\lambda Q_\lambda^T = Q_\lambda^T Q_\lambda = \openone_{2 g} \,
\ \ \ \ Q_\lambda J_{2g} Q_\lambda^{-1} = J_{2g}  \, ,
\end{equation}
indicating that $Q_\lambda$ is  an orthogonal  symplectic
transformation. Hence, $\widetilde{K}$ can be brought to the block
diagonal  form $\widetilde{K}_\lambda' = \kappa_\lambda \beta $ by
means of a one-sided orthogonal transformation $O_A =
Q_\lambda^T$, leaving the upper diagonal block of \rref{degcm}
intact. Grouping the resulting modes in each degenerate space as
$\widetilde{\gamma}_\lambda =
\oplus_{i=1}^g(\widetilde{\gamma}\subat{i}\oplus\widetilde{\gamma}\subbt{i})$,
we achieve the pair-wise decomposition for $\lambda \neq
\lambda_0$, with each pair described by a covariance matrix of the
form \rref{isoform2}. For $\lambda =\lambda_0$, the degeneracies
on each side are not restricted, and equations \rref{rel2} imply
that $\widetilde{K}_{\lambda_0} \widetilde{K}_{\lambda_0}^T =0
\Rightarrow \widetilde{K}_{\lambda_0}=0$. Therefore,  local modes
with symplectic eigenvalue $\lambda_0$ decouple, as may be
expected for the pure case $\lambda_0 = 1$.

We briefly comment on the quantification of bipartite entanglement
as obtained from the modewsie decomposition. In a fundamental
fermion theory, the characterization of entanglement is somewhat
delicate, as one needs to  take into account the absence of a
natural tensor-product decomposition of the fermionic Hilbert
space, as well as restrictions on the set of possible local
operations posed by the anticommutativity of fermion operators
\cite{BK} and  superselection rules due to fermion number
conservation\cite{WV}. Two measures are available: the explicit
entanglement content of the state  with respect to the second
quantized mode Fock space--quantified by the ``entanglememt of
modes" $E_M $~\cite{zanardi}, and the  operationally accesible
bipartite entanglemement of the state--quantified by the
``entanglement of particles" $E_P$, with $E_P\leq E_M$~\cite{WV}.
For the pure case of Eq.\rref{decomp}, $E_M$ will simply be the
sum of the von Neumann entropies of the partial density matrices
obtained from the Schmidt coefficients of the individual entangled
mode pairs in Eq.\rref{tmss}. On the other hand,  a
characterization in terms of $E_P$ is less direct, as the measure
is superadditive and will depend on the fundamental fermion
content of the mode bases in the decomposition. Nevertheless,
since $E_P$ and $E_M$ coincide asymptotically\cite{WV}, one may
expect that for entangled states involving a large number of
fermions, $E_M$ should give a reasonable indication of the usable
entanglement.

If mixed fermion states are analyzed according to the $E_M$
criterion, it is possible to asses the separability of isotropic
gaussian states via the Peres-Horodecki partial transpose (PT)
criterion \cite{peres-horodecki}. Since a two-mode mixed state
\rref{twomodemixed} is  equivalent to a two-qubit mixed state,
PT-negativity of at least one of the states
$\widetilde{\rho}\subabt{s}$ in Eq.\rref{decomp2}  will then be
sufficient~\cite{horodecki-iff} to ascertain the inseparability of
the isotropic state $\rho^{(0)}$.  Using the fact that the
entangling operator $T_i$ in \rref{ti} performs a rotation between
$\ket{00}\subabt{i}$ and $\ket{11}\subabt{i}$ while leaving
$\ket{01}\subabt{i}$ and $\ket{10}\subabt{i}$ unaltered, the
matrix representation of $\widetilde{\rho}_i$ can easily be shown
to take the block form:
\begin{equation}
[\widetilde{\rho}_i] =\frac{1}{4}\left(%
\begin{array}{cc}
  \!(1\! +\! \lambda_i)^2\! +\! \kappa_i^2 & \!2\kappa_i \\
 \! 2\kappa_i & \!(1\! -\! \lambda_i)^2\! +\! \kappa_i^2 \\
\end{array}%
\right)\oplus\left(%
\begin{array}{cc}
 \!1\!-\! \lambda_0^2 & \!0 \\
 \! 0 & \!1\! -\! \lambda_0^2 \\
\end{array}%
\right)
\end{equation}
where the first block refers to the $\ket{00}\subabt{i}$ and
$\ket{11}\subabt{i}$  vectors and the second to the
$\ket{01}\subabt{i}$ and $\ket{10}\subabt{i}$ vectors. In this
decomposition, the partial transpose operation amounts to swapping
the off-diagonal elements between the two blocks.
%\begin{equation}
%[\widetilde{\rho}_i^{PT}] =\frac{1}{4}\left(%
%\begin{array}{cc}
%  \!(1\! +\! \lambda_i)^2\! +\! \kappa_i^2 & \!0 \\
% \! 0 & \!(1\! -\! \lambda_i)^2\! +\! \kappa_i^2 \\
%\end{array}%
%\right)\oplus\left(%
%\begin{array}{cc}
% \!1\!-\! \lambda_0^2 & \! 2\kappa_i \\
% \! 2 \kappa_i & \!1\! -\! \lambda_0^2 \\
%\end{array}%
%\right)
%\end{equation}
The negative partial transpose criterion for $\widetilde{\rho}_i$
then becomes $ \lambda_0 \geq \kappa_i > \frac{1}{2}(1 -
\lambda_0^2)$. Note that this condition cannot be satisfied for
any $\kappa_i$ if $\lambda_0 < \sqrt{2}-1$ and is satisfied for
all non-zero $\kappa_i$ in the pure case $\lambda_0=1$ as
expected.

In conclusion, we have studied fermion mode entanglement and
showed that under an arbitrary bi-partite division of a set of
fermion modes, gaussian mode entanglement has a universal
structure of a generalized BCS-like wave-function. Interestingly,
unlike the gaussian boson case, our result applies not only to the
ground state, but also to all definite excitation eigenstates of
quadratic fermion Hamiltonians. Our result can be used for
computing the bi-partite mode entanglement of a variety of states
in a wide class of models, such as certain spin-chain models, and
fermion liquids and superconductivity in the mean-field
approximation, we believe that the present results can be helpful
in connection with information processing applications involving
coupled many-particle fermion systems or superconductors. For
instance, when two such systems are coupled by a quadratic
tunnelling interaction \cite{cohen}, our result implies that under
a proper choice of ``local operators", the total wave function
factorizes to a product of non-maximal EPR-like states.

A.B. acknowledges support from Colciencias (contract
1204-05-13622). B.R. acknowledges the support of ISF grant
62/01-1.

\end{document}